\documentclass[a4paper,12pt]{article}
\usepackage{amsmath,amssymb,epic,epsfig,float,cite}
\newtheorem{thm}{\bf Theorem}[section]

\newtheorem{prop}[thm]{\bf Proposition}

\numberwithin{equation}{section}

\newcommand{\al}{\alpha}
\newcommand{\de}{\delta}
\let\ep\epsilon
\newcommand{\om}{\omega}
\let\ph\phi
\let\si\sigma
\let\z\zeta

\newcommand{\cL}{{\cal L}}
\newcommand{\cP}{{\cal P}}

\newcommand{\R}{{\mathbb R}}
\newcommand{\N}{{\mathbb N}}
\newcommand{\C}{{\mathbb C}}

\def\sL{\mathfrak{sl}}

\let\d\partial
\def\bh{\hat b}
\def\ch{\hat c}
\def\dh{\hat d}
\newcommand{\csch}{\operatorname{csch}}
\newcommand{\arccosh}{\operatorname{arccosh}}
\let\union\cup

\let\Em\bf
\def\Hhg{\hat H_{\text{g}}}
\def\Hhgp{\hat H^{\text{per}}_{\text{g}}}

\def\Htgp{\tilde H^{\text{per}}_{\text{g}}}
\def\Pp{\hat P^{\text{per}}}
\def\Ptp{\tilde P^{\text{per}}}
\def\chitp{\tilde\chi^{\text{per}}_E}
\def\ptp{\tilde\psi_E^{\text{per}}}
\def\ptpm{\tilde\psi_{-E}^{\text{per}}}
\def\pb{\bar\psi}
\def\pp{\tilde\psi^{\text{per}}}

\newcommand{\fdim}{fi\-nite-di\-men\-sion\-al}
\newcommand{\QED}{\quad Q.E.D.}
\def\ie{i.e.}
\begin{document}
\title{\null\vskip-1cm\textbf{\Large On the Families of Orthogonal 
Polynomials\\
Associated to the Razavy Potential}%
\thanks{Supported in part by DGES Grant PB95--0401.}\vskip.5cm}
\author{\sc Federico Finkel\footnote{On
leave of absence from Depto.~de F\'\i sica Te\'orica II,
Univ.~Complutense de Madrid, Spain.}\\
\em Department of Mathematics\\
\em Imperial College\\
\em London SW7 2BZ, UK\\
\\
\sc Artemio Gonz\'alez-L\'opez, Miguel A.~Rodr\'\i guez\\
\em Departamento de F\'\i sica Te\'orica II\\
\em Universidad Complutense de Madrid\\
\em 28040 Madrid, SPAIN\\[.7cm]}
\date{May 11, 1999; revised July 30, 1999}
\maketitle
\vskip5cm
\noindent
Short Title:\quad{\em Orthogonal Polynomials and the Razavy Potential}\\[1mm]
\newpage
\begin{abstract}
We show that there are two different families of (weakly) orthogonal
polynomials associated to the quasi-exactly solvable
Razavy potential $V(x)=(\z\,\cosh 2x-M)^2$ ($\z>0$,
$M\in\N$). One of these families encompasses the four sets of
orthogonal polynomials recently found by Khare and Mandal, while the
other one is new. These results are extended to the related periodic
potential $U(x)=-(\z\,\cos 2x -M)^2$, for which we also construct two
different families of weakly orthogonal polynomials. We prove that
either of these two families yields the ground state (when $M$ is odd)
and the lowest lying gaps in the energy spectrum of the latter
periodic potential up to and including the $(M-1)^{\rm th}$ gap and
having the same parity as $M-1$. Moreover, we show that the algebraic
eigenfunctions obtained in this way are the well-known finite solutions
of the Whittaker--Hill (or Hill's three-term) periodic differential
equation. Thus, the foregoing results provide a Lie-algebraic
justification of the fact that the Whittaker--Hill equation (unlike,
for instance, Mathieu's equation) admits finite solutions.
\end{abstract}
\newpage
\section{Introduction}
\label{sec.intro}
The one-dimensional quantum mechanical potential
\begin{equation}
\label{razavy}
V(x)=(\z\,\cosh 2x-M)^2,
\end{equation}
where $\z$ and $M$ are positive real parameters, was first studied by
Razavy \cite{Ra80}. For $M>\z$, the above potential (to which we shall
henceforth refer as the {\Em Razavy potential}) is a symmetric double
well. This type of potentials has been extensively used in the quantum
theory of molecules as an approximate description of the motion of a
particle under two centers of force. In particular, the Razavy
potential has been proposed by several authors as a realistic model of a proton
in a hydrogen bond, \cite{LR81, RL81, MM82, DS92}. The potential
\eqref{razavy} has also been recently used by Ulyanov and Zaslavskii,
\cite{UZ92}, as an effective potential for a uniaxial paramagnet.

Razavy showed that when $M$ is a positive integer the lowest $M$
energy levels of the potential \eqref{razavy} (with their
corresponding eigenfunctions) can be exactly computed in closed form.
The Razavy potential is thus an example of a {\Em quasi-exactly
solvable} (QES) potential, for which part (but not necessarily all) of
the spectrum can be computed exactly. A very important class of QES
potentials, that we shall call {\Em algebraic} in what
follows\footnote{There is, unfortunately, no clear consensus in the
literature regarding this terminology. The term ``quasi-exactly
solvable potential" is, we believe, originally due to Turbiner and
Ushveridze, \cite{TU87}, who used it to refer to what we have just
called {\em algebraic} QES potentials. However, in the last couple of
years there has been a growing tendency to use the adjective
``quasi-exactly solvable" for any potential, be it algebraic or not,
part of whose spectrum can be exactly computed. We have preferred in
this paper to adhere to this increasingly more common usage to avoid
confusion.}, are characterized by the fact that the corresponding
quantum Hamiltonian is an element of the enveloping algebra of a
\fdim{} Lie algebra of differential operators (the so-called {\Em
hidden symmetry algebra}) admitting a finite-dimensional invariant
module of smooth functions. That such a potential is QES follows
immediately from the fact that the finite-dimensional module of the
hidden symmetry algebra is obviously left invariant by its enveloping
algebra, and in particular by the Hamiltonian. Therefore, a number of
energy eigenvalues and eigenfunctions equal to the dimension of the
invariant module can be computed algebraically, by diagonalizing the
finite-dimensional matrix of the restriction of the Hamiltonian to the
module.

One-dimensional algebraic QES potentials were studied as such for the first
time by Turbiner \cite{Tu88}, who used as hidden symmetry algebra a
realization of $\sL(2,\R)$ in terms of first-order differential operators.
These potentials were then completely classified by Gonz\'alez-L\'opez,
Kamran and Olver \cite{GKOnorm93, GKOqes94}. There are exactly ten
families of one-dimensional algebraic QES potentials, five of which are
periodic and the remaining five all have point spectrum. In all cases, the
hidden symmetry algebra is again $\sL(2,\R)$.

Recently Bender and Dunne \cite{BD96} associated a family of
(weakly) orthogonal polynomials to the class of algebraic QES potentials
given by
$$
V(x) = \frac{(4s-1)(4s-3)}{4x^2}-2(2s+2J-1)x^2+x^6;
\qquad s\in\R,\quad J\in\N.
$$
This construction was immediately extended by the authors of this paper to
virtually all one-dimensional {\em algebraic} QES potentials in
\cite{FGRorth96}. Krajewska, Ushveridze and Walczak \cite{KUW97}
proved that a set of weakly orthogonal polynomials can be constructed
explicitly for any (not necessarily algebraic) QES Hamiltonian
tridiagonalizable in a {\em known} basis. Khare and Mandal
have constructed two families of non-orthogonal polynomials associated
to a pair of non-algebraic QES potentials,
\cite{KM98a}\footnote{Some of the formulae for the polynomials associated
to these potentials contain errata.
Indeed, the change of variable (8) should read $t=(y+\ep^2)^{1/2}$,
and the factor multiplying $n$ in the coefficient of $Q_n(s)$ in the
recursion
relation (11) should be $\ep^2(4s+1)$, this affecting formulae (12) and
(13).
Likewise, in formula (21) the term $(a+b+c-\frac32)^2$ should
be $(a+b+c+n-\frac32)^2$.}.
It is important to note that the family of
polynomials associated to a given Hamiltonian is not unique, but depends
on the type of formal expansion defining the polynomials.
It is therefore conceivable that one could
obtain orthogonal polynomials in the examples studied in \cite{KM98a} by
considering different expansions.

The Razavy potential has been recently revisited by Khare and Mandal,
\cite{KM98}, and Konwent et al., \cite{KMMR98}. The former authors, who
were mainly interested in the properties of the associated polynomial
system, introduced four different sets of polynomials
$\{P^\ep_k(E)\}_{k=0}^\infty$ and $\{Q^\ep_k(E)\}_{k=0}^\infty$
($\ep=0,1$) for the Razavy potential \eqref{razavy} through the
formulae
$$
\psi_E(x) = e^{-\frac\z2 z} (z-1)^{\ep/2} \sum_{k=0}^\infty
\frac{P^\ep_k(E)}{(2k)!}
\left(\frac{z+1}2\right)^k
$$
and
$$
\psi_E(x) = e^{-\frac\z2 z} (z-1)^{\ep/2} \sum_{k=0}^\infty
\frac{Q^\ep_k(E)}{(2k+1)!}
\left(\frac{z+1}2\right)^{k+\frac12},
$$
where
\begin{equation}
\label{z}
z=\cosh 2x,
\end{equation}
and $\psi_E$ denotes a formal (i.e., not necessarily square-integrable)
eigenfunction of
$$
H=-\d_x^2+(\z\,\cosh 2x-M)^2
$$
with eigenvalue $E$ and parity\footnote{Where $(z-1)^{1/2}$ should of
course be interpreted as $\sqrt 2\sinh x$.} $(-1)^\ep$. Without loss
of generality, we shall choose the usual normalization
$$
P^\ep_0(E)=Q^\ep_0(E)=1.
$$
Imposing that $(H-E)\,\psi_E(x)=0$ one easily shows that each of the
four sets $\{P^\ep_k(E)\}_{k=0}^\infty$ and
$\{Q^\ep_k(E)\}_{k=0}^\infty$ ($\ep=0,1$) satisfies a three-term recurrence
relation of the appropriate form (see~\eqref{rr} and Ref.~\cite{Ch78}),
and therefore forms an orthogonal
polynomial system with respect to a suitable Stieltjes measure.

We thus have four seemingly unrelated sets of orthogonal polynomials
associated to the Razavy potential
\eqref{razavy}. This is surprising, since in all the previous examples
only one set of orthogonal polynomials was constructed for each QES
potential considered. One of the objectives of this paper is precisely to
explain how these four sets of orthogonal polynomials arise. The key
to this explanation is the fact (not taken into account in \cite{KM98})
that the Razavy potential is not just QES, but {\em algebraic}. More
precisely, we shall show in Section \ref{sec.razavy} that the Razavy
potential can be written as a polynomial in the generators of a suitable
realization of $\sL(2,\R)$ in two different ways. Using the constructive
method explained in \cite{FGRorth96}, these two different realizations of
the Razavy potential as an algebraic QES potential give rise to two
different families of orthogonal polynomials. One of these two families
encompasses in a natural way the four sets of orthogonal polynomials
of Khare and Mandal's. In fact, all the properties of these four sets
found in \cite{KM98} (weak orthogonality, factorization, etc.) are
immediate consequences of the general properties of the system of
orthogonal polynomials associated to an algebraic QES potential developed
in our previous paper \cite{FGRorth96}. The second realization of
\eqref{razavy} as an algebraic QES potential yields yet another set of
orthogonal polynomials different from the four sets found by Khare and
Mandal. The properties of this family, which again follow from the
general theory developed in \cite{FGRorth96}, are in many respects simpler
than those of the first family. For example, the moment functional
associated to the second family is positive semidefinite, while this is not
the case for the first family. All of these facts make, in our opinion, the
second family more convenient in practice for finding
the exactly computable energy levels of the Razavy
potential.

In Section \ref{sec.trazavy} we study the trigonometric version of
the Razavy potential, given by
\begin{equation}
\label{trazavy}
U(x)=-(\z\,\cos 2x - M)^2.
\end{equation}
This potential, which is a simple model for a one-dimensional periodic
lattice, appears in Turbiner's list of QES one-dimensional potentials,
\cite{Tu88}, and was also touched upon by Shifman, \cite{Sh89} (in the
particular case in which $M$ is an odd positive integer). Ulyanov and
Zaslavskii, \cite{UZ92}, have related the trigonometric Razavy
potential \eqref{trazavy} to a quantum spin system. The potential
\eqref{trazavy} has also recently appeared as the coupling term
between the inflaton field and matter scalar fields in theories of
cosmological reheating after inflation with a displaced harmonic
inflaton potential, \cite{KLS97}.

\begin{figure}[h]
\begin{center}
\begin{picture}(400,50)(-200,-20)
\put(-200,18){\small $M=4$}
\dottedline{2}(-150,20)(-140,20)
\put(-110,20){\line(1,0){30}}
\dottedline{2}(-50,20)(-20,20)
\put(10,20){\line(1,0){30}}
\dottedline{2}(70,20)(100,20)
\dottedline{2}(130,20)(160,20)
\thicklines
\put(-110,20){\circle*{5}}\put(-140,19.75){\framebox(30,.5)}
\put(-140,20){\line(1,0){30}}
\put(-80,20){\circle*{5}}\put(-80,19.75){\framebox(30,.5)}
\put(-80,20){\line(1,0){30}}
\put(10,20){\circle*{5}}\put(-20,19.75){\framebox(30,.5)}
\put(-20,20){\line(1,0){30}}
\put(40,20){\circle*{5}}\put(40,19.75){\framebox(30,.5)}
\put(40,20){\line(1,0){30}}
\put(100,19.75){\framebox(30,.5)}
\put(100,20){\line(1,0){30}}
\thinlines
\put(-200,-22){\small $M=5$}
\put(-150,-20){\line(1,0){10}}
\dottedline{2}(-110,-20)(-80,-20)
\put(-50,-20){\line(1,0){30}}
\dottedline{2}(10,-20)(40,-20)
\put(70,-20){\line(1,0){30}}
\dottedline{2}(130,-20)(160,-20)
\thicklines
\put(-140,-20){\circle*{5}}\put(-140,-20.25){\framebox(30,.5)}
\put(-140,-20){\line(1,0){30}}
\put(-50,-20){\circle*{5}}\put(-80,-20.25){\framebox(30,.5)}
\put(-80,-20){\line(1,0){30}}
\put(-20,-20){\circle*{5}}\put(-20,-20.25){\framebox(30,.5)}
\put(-20,-20){\line(1,0){30}}
\put(70,-20){\circle*{5}}\put(40,-20.25){\framebox(30,.5)}
\put(40,-20){\line(1,0){30}}
\put(100,-20){\circle*{5}}\put(100,-20.25){\framebox(30,.5)}
\put(100,-20){\line(1,0){30}}
\end{picture}
\end{center}
\begin{quote}
\caption{Structure of the energy spectrum of the potential
\eqref{trazavy} for $M=4$ and $M=5$. The thick horizontal segments
represent the allowed energy bands. The solid circles stand for the
algebraically computable energies, which determine the gaps 
represented by thin solid lines. The remaining boundaries of
the allowed bands, and hence the energy gaps shown as dotted lines,
cannot be exactly computed.}\label{fig1}
\end{quote}
\end{figure}
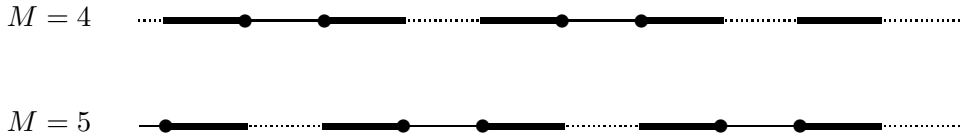

The trigonometric Razavy potential \eqref{trazavy} is the image of the
hyperbolic Razavy potential \eqref{razavy} under the anti-isospectral
transformation $x\mapsto ix$, $E\mapsto -E$, recently considered by
Krajewska, Ushveridze and Walczak \cite{KUWdual97}. It is therefore to be
expected that the properties of the polynomials associated to this
potential are analogous to the corresponding properties for the hyperbolic
Razavy potential \eqref{razavy}. That this is indeed the case is shown in
Section~\ref{sec.trazavy}, where we prove that the potential
\eqref{trazavy} can be realized in two different ways as an algebraic QES
potential. As in the hyperbolic case, each of these two different
realizations gives rise to a family of orthogonal polynomials. For each
positive integer value of $M$ it is possible to exactly compute $M$
eigenfunctions (with their corresponding energies) of the trigonometric
Razavy potential by purely algebraic procedures. It was previously known on
general grounds \cite{Sh89} that the energies of these $M$ eigenfunctions
must be boundary points of allowed bands (or, equivalently, gaps) in the
energy spectrum of the periodic potential \eqref{trazavy}. In Section
\ref{sec.trazavy} we investigate the exact position of these boundary
points in the spectrum of \eqref{trazavy}. We will show that, if the gaps
in the energy spectrum are numbered consecutively in order of increasing
energy, these points yield precisely the ground state (when $M$ is odd)
and the lowest $\left[\frac M2\right]$ gaps\footnote{We denote by $[x]$ the
integer part of the real number $x$.} of the same parity as
$M-1$. For instance, if $M=4$ we obtain the first (lowest) and the third
gap, whereas for $M=5$ we get the ground state and the second and fourth
lowest gaps (see Fig.~\ref{fig1}).

The paper ends with a discussion of the above results in the context
of the classical theory of Hill's equation. We show that the
algebraic eigenfunctions constructed in this paper are precisely the
so called finite solutions of the Whittaker--Hill (or Hill's three-term)
equation. In fact, our analysis provides a Lie-algebraic explanation
of why the Whittaker--Hill equation admits finite solutions at all.
Indeed, from our point of view this is just a simple consequence of
the fact that the Schr\"odinger operator with potential \eqref{trazavy} is
algebraically QES.
\section{The hyperbolic Razavy potential}
\label{sec.razavy}
We shall show in this section that the hyperbolic Razavy potential
\eqref{razavy} can be expressed in two different ways as an algebraic QES
potential. From these two representations we shall derive two
different families of associated orthogonal polynomials, whose properties
we shall discuss.

Consider, in the first place, the second non-periodic algebraic QES
potential listed in Ref.~\cite{GKOqes94} (p.~127), given by
\begin{equation}\label{case2}
V(x) = A\,\cosh^2\sqrt\nu\,x + B\,\cosh\sqrt\nu\,x
+ C\,\coth\sqrt\nu\,x\,\csch\sqrt\nu\,x
+ D\,\csch^2\sqrt\nu\,x\,,
\end{equation}
where the coefficients $A,B,C,D$ can be
expressed in terms of four parameters $\bh,\ch,\dh\in\R$ and
$n\in\N\union\{0\}$ as follows (see \cite{GKOnorm93}, Eq.~(5.11)):
\begin{gather}
A = \frac{\bh^2}{4\nu},\qquad B = \frac
\bh{2\nu}\left[\ch+(n+1)\nu\right],
\qquad C=\frac{\bh+\dh}{2\nu}\left[\ch-(n+1)\nu\right]\notag\\
\label{param}
D=\frac1{4\nu}\left[(\bh+\dh)^2+\bigl(\ch-(n+1)\nu\bigr)^2
-\nu^2\right].
\end{gather}

% QES form in terms of the parameters. Normalizibility conditions.
The hyperbolic Razavy potential is of the form \eqref{case2} (up
to an inessential additive constant) provided that
$$
\nu=4,\qquad C=D=0,\qquad A=\z^2,\qquad B=-2M\z.
$$
\goodbreak Using Eq.~\eqref{param} we obtain the following system
in the parameters $\bh$, $\ch$, $\dh$ and $n$:
\begin{align}
\bh^2&=16\,\z^2,\label{sys1}
\\
\bh\,\left[\ch+4(n+1)\right]&=-16\,M\,\z,\label{sys2}
\\
(\bh+\dh)\left[\ch-4(n+1)\right]&=0,\label{sys3}
\\
(\bh+\dh)^2+\left[\ch-4(n+1)\right]^2\label{sys4}
&=16\,.
\end{align}
From Eqs.~\eqref{sys1} and~\eqref{sys2} and the normalizability
condition $\bh<0$ (see~\cite{GKOnorm93}, Eq.~(5.14)), we get
$$
\bh=-4\z\,,\qquad \ch=4(M-n-1)\,.
$$
Substituting these into the remaining conditions
\eqref{sys3} and \eqref{sys4} we are led to four different
solutions for $\dh$ and $M$, which may be written in a unified way as
$$
M=2(n+1)-\si\,,\qquad \ch=4(n+1-\si)\,,\qquad \dh=4(\z-\eta)\,,
$$
where the parameters $\si$ and $\eta$ are given in Table~\ref{table1}.
\smallskip
\begin{table}
\begin{center}
\begin{tabular}{|c|c||r|r|}
\hline
$\dh$ & $M$ & $\si$ & $\eta$\\
\hline
$4\z$ & $2n+1$ & $1$ & $0$\\
$4\z$ & $2n+3$ & $-1$ & $0$\\
$4(\z-1)$ & $2n+2$ & $0$ & $1$ \\
$4(\z+1)$ & $2n+2$ & $0$ & $-1$\\
\hline
\end{tabular}
\begin{quote}
\caption{Values of $\hat d$ and $M$ corresponding to the four 
solutions of Eqs.~\eqref{sys1}--\eqref{sys4}.}\label{table1}
\end{quote}
\end{center}
\end{table}
%\smallskip
\smallskip

It follows from the general discussion in Ref.~\cite{GKOnorm93}
that the change of variable~\eqref{z} and the gauge transformation
determined by
\begin{equation}\label{mu2}
\hat\mu(z)=(z-1)^{\frac14(1-\si-\eta)}(z+1)^{\frac14(1-\si+\eta)}e^{-\z z/2}
\end{equation}
map the Razavy Hamiltonian into an operator $\hat H_{\text{g}}$
(the {\bf gauge Hamiltonian})
belonging to the enveloping algebra of the realization of $\sL(2,\R)$
spanned by\footnote{The operators $J_\al$ ($\al=\pm,0$)
(and any polynomial thereof) preserve the space ${\cal P}_n$ of
polynomials in $z$ of degree at most $n$. Moreover, any
$k$-th order differential operator ($k\leq n$) preserving ${\cal P}_n$
may be expressed as a $k$-th degree polynomial in the
generators $J_\al$, \cite{Tu92, FK98}.}
\begin{equation}\label{Js}
J_-=\d_z\,,\qquad J_0=z\d_z-\frac n2\,,\qquad J_+=z^2\d_z-n\,z\,.
\end{equation}
Indeed,
\begin{equation}\label{Hgauge}
\begin{aligned}
\hat H_{\text{g}}(z)&=\left.\frac1{\hat\mu(z)}\cdot
\left[-\d_x^2+(\z\cosh 2x-M)^2\right]
\cdot\hat\mu(z)\right|_{x=\frac12\arccosh z}\\
&=-4 \left(J_0^2-J_-^2-\z J_++(n+1-\si)J_0+(\z-\eta)J_-\right)-c_\ast\,,
\end{aligned}
\end{equation}
where
$$
c_\ast=-3(n+1)^2+2(n+1)\si+2\eta\z-\z^2.
$$
According to the general prescription of Ref.~\cite{FGRorth96},
the formal solutions of the gauged equation
\begin{equation}\label{gaugedeq}
(\hat H_{\text{g}}-E)\hat\chi_E=0\,,
\end{equation}
are generating functions for a set of orthogonal polynomials.
More explicitly, inserting the expansion
\begin{equation}\label{series2}
\hat\chi_E(z)=\sum_{k=0}^\infty
\frac{\hat P_k(E)}{2^{k}\,\big(2k+\frac{\eta-\si+1}2\big)!}\,(z+1)^k\,,
\end{equation}
into~\eqref{gaugedeq}, we readily find that the coefficients $\hat P_k(E)$
satisfy the three-term recursion relation
\begin{equation}\label{rr}
\hat P_{k+1}(E) = (E-b_k)\,\hat P_k(E)-a_k\,\hat P_{k-1}(E),\qquad k\geq 0,
\end{equation}
where
\begin{equation}\label{abKM}
\begin{aligned}
a_k &=16\zeta k(2k-\sigma+\eta)(k-n-1)\,,\\
b_k &=-4k(k+1-\sigma+2\zeta)+(2n+1)(2(n-\sigma)+3)+\zeta(\zeta-2\eta+4n)\,.
\end{aligned}
\end{equation}
If we impose the condition $\hat P_0(E)=1$, the coefficients $\{\hat
P_k(E)\}_{k=0}^\infty$
form a set of weakly orthogonal (monic) polynomials. Therefore,
we can construct two sets of weakly orthogonal
polynomials for each value of $M$ by choosing suitable values for
$\si$, $\eta$ and $n$ (for $M=1$ there is only one set).
These two sets coincide exactly with the sets
$\{P^{1-\ep}_k(E)\}_{k=0}^\infty$
and $\{Q^{\ep}_k(E)\}_{k=0}^\infty$ (with $\ep=0$ for $M$ even and
$\ep=1$ for $M$ odd) studied by Khare and Mandal in~\cite{KM98}.
From now on, we shall use when necessary the more precise notation
$\hat P^{\si\eta}_k$ to denote the orthogonal polynomials defined by
Eqs.~\eqref{rr}--\eqref{abKM}.
For instance, if $M=4$ we have $n=1$, $\si=0$
and $\eta=\pm 1$ (see Table~\ref{table1}). When $\eta=1$ the first three polynomials are
\begin{equation}\label{etap1}
\begin{aligned}
\hat{P}^{0+}_0(E) &=1,\\
\hat{P}^{0+}_1(E) &= E-\zeta^2-2\zeta-15,\\
\hat{P}^{0+}_2(E) &= E^2-2(\zeta^2-2\zeta+11)E+\zeta^4-
4\zeta^3+10\zeta^2-28\zeta+105,
\end{aligned}
\end{equation}
whereas for $\eta=-1$ we obtain
\begin{equation}\label{etam1}
\begin{aligned}
\hat{P}^{0-}_0(E) &=1,\\
\hat{P}^{0-}_1(E) &= E-\zeta^2-6\zeta-15,\\
\hat{P}^{0-}_2(E) &= E^2-2(\zeta^2+2\zeta+11)E+\zeta^4+
4\zeta^3+10\zeta^2+28\zeta+105.
\end{aligned}
\end{equation}
The polynomials~\eqref{etap1}
and~\eqref{etam1} reduce, respectively,
to the polynomials $Q^0_k(E)$ and $P^1_k(E)$ ($k=0,1,2$)
in formulae (2.18) and (2.17) of Ref.~\cite{KM98}.

The coefficient $a_k$ given by~\eqref{abKM} vanishes for $k=n+1$, and
therefore $\hat P_k(E)$ with $k\geq n+1$ factorize as $\hat
P_{n+1+j}(E)=\hat Q_j(E)\hat P_{n+1}(E)$, $j\geq 0$, where $\{\hat
Q_j\}_{j=0}^\infty$ also form a set of (monic) orthogonal polynomials.
If $E_j$ is a root of the polynomial $\hat P_{n+1}(E)$ the
series~\eqref{series2} truncates at $k=n$, and thus $E_j$ belongs to
the point spectrum of the Razavy Hamiltonian. For example, if $M=4$,
the roots $E_0$, $E_2$ of $\hat{P}^{0+}_2(E)$ are the energies of the ground
state and the second excited state of the Razavy potential, while the
roots $E_1$, $E_3$ of $\hat{P}^{0-}_2(E)$ correspond to the first and third
excited levels. The rest of the spectrum cannot be computed
algebraically.

The other usual properties which characterize weak
orthogonality ---vanishing norms, finite support of the
Stieltjes measure associated to the polynomials,
etc.,~\cite{BD96, FGRorth96, KUW97}---
are also satisfied by the polynomials $\{\hat P_k(E)\}_{k=0}^\infty$.
In particular, if $E_k$ ($k=0,\dots,n$) are the (different)
roots of $\hat P_{n+1}(E)$, the moment functional associated to
the polynomials is
\begin{equation}\label{mf}
\cL=\sum_{k=0}^n \om_k\de(E-E_k),
\end{equation}
where the coefficients $\om_k$ are determined by
$$
\sum_{k=0}^n{\hat P}_l(E_k)\om_k=\de_{l0},\qquad l=0,\dots,n.
$$
It was observed in~\cite{KM98} that not all the coefficients $\om_k$
corresponding to the polynomials $P^\ep_k$ and $Q^\ep_k$
are positive. This is in fact a direct consequence of the following
general property of an orthogonal polynomial system satisfying a recursion
relation of the form~\eqref{rr}.
\begin{prop}\label{prop1}
The coefficients $\om_k$ of the moment functional~\eqref{mf}
are positive for all $k=0,\dots,n$ if and only if
$a_k>0$ for $0<k\leq n$ and $b_k$ is real for $0\leq k<n$.
\end{prop}
{\em Proof.} The ``if\kern1pt" part was proved
in~\cite{FGRorth96}\footnote{Note that
the proof only requires $b_k$ to be real for $0\leq k<n$.}. Let $\om_k>0$
for $k=0,\dots,n$. Then $\cL(p^2)>0$ for any non-vanishing
real polynomial of degree at most $n$. It follows that $\cL$
is positive definite in $\cP_{2n}=\{P\in\C[E]:\deg P\leq 2n\}$,
for if $P\in\cP_{2n}$ is a nonzero real polynomial which is non-negative for
all $E\in\R$, then $P=p^2+q^2$ for real polynomials $p$ and $q$ in $\cP_n$,
and thus $\cL(P)>0$. Since $\cL$ is positive definite in $\cP_{2n}$,
the moments $\mu_k=\cL(E^k)$ with $k=0,\dots,2n$ are positive for even $k$
and real for odd $k$, \cite{Ch78}. Multiplying the recursion
relation~\eqref{rr} by $\hat P_k$ and applying $\cL$ we find that
\begin{equation}\label{proof}
\cL(E{\hat P}^2_k)-b_k\cL({\hat P}^2_k)=0\,.
\end{equation}
Taking $k=0$, we conclude that $b_0=\mu_1/\mu_0$ is real. Therefore
$\hat P_1=E-b_0$ is real and $a_1=\cL(\hat P_1^2)>0$. By induction, if
$b_{j-1}\in\R$ and $a_j>0$ for $j=1,\dots,k<n$, then $\hat P_k$
is real, and from~\eqref{proof} we deduce that $b_k\in\R$. Then
$\hat P_{k+1}$ is real, and
$$
0<\cL(\hat P_{k+1}^2)=\prod_{j=1}^{k+1}a_j
$$
implies that $a_{k+1}>0$.\QED\\

Note that the coefficients $a_k$ given by~\eqref{abKM} are negative for
$1\leq k\leq n$ and therefore $\om_k$ cannot be positive for all
$k=0,\dots,n$.\\

Consider, in the second place, the third non-periodic algebraic
QES potential given in Ref.~\cite{GKOqes94} (p.~127), namely
\begin{equation}\label{case3}
V(x) = A\,e^{2\sqrt\nu\,x} + B\,e^{\sqrt\nu\,x}
+ C\,e^{-\sqrt\nu\,x}+D\,e^{-2\sqrt\nu\,x}\,,
\end{equation}
where the coefficients $A,B,C,D$ can again be
expressed in terms of four parameters $\bh,\ch,\dh\in\R$ and
$n\in\N\union\{0\}$ as (see \cite{GKOnorm93}, Eq.~(5.11)):
\begin{equation}\label{param3}
A=\frac{\bh^2}{4\nu},\qquad B=\frac\bh{2\nu}\left[\ch+(n+1)\nu\right],\qquad
C=\frac\dh{2\nu}\left[\ch-(n+1)\nu\right],\qquad D=\frac{\dh^2}{4\nu}\,.
\end{equation}
The potential~\eqref{case3} reduces to the hyperbolic Razavy
potential~\eqref{razavy} (up to an additive constant)
provided that
$$
\nu=4,\qquad A=D=\frac{\z^2}4,\qquad B=C=-\z M.
$$
Taking into account the normalizability conditions $\bh<0$ and $\dh>0$,
(see~\cite{GKOnorm93}, Eqs.~(5.24) and (5.25)), we get the unique
solution
$$
\bh=-2\z,\qquad \ch=0,\qquad \dh=2\z,\qquad M=n+1.
$$
In this case, the change of variable
\begin{equation}\label{z3}
z=e^{2x}
\end{equation}
and the gauge transformation generated by
\begin{equation}\label{mu3}
\tilde\mu(z)=z^{\frac{1-M}2}e^{-\frac\z4(z+\frac1z)}
\end{equation}
map the Razavy Hamiltonian into a differential
operator $\tilde H_{\text{g}}$ quadratic in the
generators~\eqref{Js}, namely
\begin{equation}\label{tHgauge}
\tilde H_{\text{g}}(z)=-4 J_0^2+2\z J_+-2\z J_--\tilde c_\ast\,,
\end{equation}
where
$$
\tilde c_\ast=-(n+1)^2-\z^2.
$$
Following the general treatment of~\cite{FGRorth96}, if we insert the
expansion
\begin{equation}\label{series3}
\tilde\chi_E(z)=\sum_{k=0}^\infty
\frac{(-1)^k\tilde P_k(E)}{(2\z)^k k!}\,z^k
\end{equation}
into the spectral equation for $\tilde H_{\text{g}}$, the
coefficients $\tilde P_k(E)$ are easily found to satisfy a three-term
recursion relation of the form~\eqref{rr}, with coefficients
\begin{equation}\label{ab3}
\begin{aligned}
a_k &=4k(n+1-k)\z^2\,,\\
b_k &=4k(n-k)+2n+1+\z^2\,.
\end{aligned}
\end{equation}
Taking $\tilde P_0(E)=1$, we obtain yet another family of weakly orthogonal
(monic) polynomials $\{\tilde P_k(E)\}_{k=0}^\infty$ associated to
the Razavy potential~\eqref{razavy}. For instance, if $M=4$ the
first five polynomials are
\begin{equation}\label{pol3}
\begin{aligned}
\tilde P_0(E)&=1,\\
\tilde P_1(E)&=E-\z^2-7,\\
\tilde P_2(E)&=E^2-2(\z^2+11)E+\z^4+10\z^2+105,\\
\tilde
P_3(E)&=E^3-(3\z^2+37)E^2+(3\z^4+46\z^2+435)E-\z^6-9\z^4-143\z^2-1575,\\
\tilde
P_4(E)&=E^4-4(\z^2+11)E^3+2(3\z^4+46\z^2+347)E^2-4(\z^6+13\z^4+159\z^2\\
&\quad +1155)E+\z^8+4\z^6+86\z^4+1316\z^2+11025.
\end{aligned}
\end{equation}
Note that the polynomial $\tilde P_4(E)$ is the product of the
polynomials $\hat P^{0+}_2(E)$ and $\hat P^{0-}_2(E)$ given in~\eqref{etap1}
and~\eqref{etam1}. Therefore, the algebraic levels can be also
obtained as the the roots $E_0,\dots,E_3$ of $\tilde P_4(E)$.

In general, if $M$ is even the polynomial $\tilde P_M(E)$ factorizes
into the product of the polynomials $\hat P^{0\pm}_{M/2}(E)$
associated to $\eta=\pm 1$ (see Table~\ref{table1}). Alternatively, if
$M$ is odd, $\tilde P_M(E)$ factorizes into the product of
the polynomials $\hat P^{\pm0}_{(M\pm 1)/2}$ associated to $\si=\pm 1$.
The algebraic energy levels of the Razavy potential~\eqref{razavy} can
thus be computed in a unified way as the roots of $\tilde P_M$.
On the other hand, the algebraic eigenfunctions can be written as
$$
\psi_E(x)=\left.\mu(z)\chi_E(z)\right|_{z=z(x)},
$$
where $\mu(z)$, $\chi_E(z)$ and the change of variable $z=z(x)$
are given by either~\eqref{z},~\eqref{mu2} and~\eqref{series2},
or~\eqref{z3},~\eqref{mu3} and~\eqref{series3}.

The polynomials $\{\tilde P_k\}_{k=0}^\infty$ verify the usual
properties associated to their weak orthogonality. However,
unlike the previous family $\{\hat P_k\}_{k=0}^\infty$,
the coefficients $\tilde a_k$ of the recursion
relation are positive for $0<k\leq n$. It follows from
Proposition~\ref{prop1} that
the coefficients $\om_k$ of the corresponding moment
functional $\cL$ are positive for all $k=0,\dots,n$,
\ie, $\cL$ is positive semidefinite.

Before finishing this section, let us emphasize that the Razavy
Hamiltonian admits two different gauged forms $\hat H_{\text{g}}$
and $\tilde H_{\text{g}}$ inequivalent under the action of the
projective group on the enveloping algebra of the
generators~\eqref{Js}, \cite{GKOnorm93}. This does not contradict the
fact that $\hat H_{\text{g}}$ and $\tilde H_{\text{g}}$ are
equivalent under a change of variable and a gauge transformation
(since they are both equivalent to the Razavy Hamiltonian). Indeed,
the transformation relating $\hat H_{\text{g}}$ and $\tilde
H_{\text{g}}$,
$$
\hat H_{\text{g}}(\hat z)=
\left.\left(\frac{\tilde\mu(\tilde z)}{\hat\mu(\hat z)}\right)
\tilde H_{\text{g}}(\tilde z)
\left(\frac{\tilde\mu(\tilde z)}{\hat\mu(\hat z)}\right)^{\!\!-1}\,
\right|_{\hat z=\frac12(\tilde z+\tilde z^{-1})},
$$
is certainly {\em not} projective.
\section{The trigonometric Razavy potential}
\label{sec.trazavy}
\subsection{The orthogonal polynomial families}
\label{sub.opf}
We shall study in this section the {\bf trigonometric Razavy potential}
\begin{equation}
  \label{tr}
U(x)=-(\z\,\cos 2x - M)^2\,,
\end{equation}
which can be obtained from the hyperbolic Razavy potential
\eqref{razavy} applying the anti-isospectral transformation $x\mapsto
i\,x$, $E\mapsto-E$. In other words, $\psi(x)$ is a solution of the
differential equation
\begin{equation}
  \label{hv}
\left[-\d_x^2+V(x)\right]\,\psi(x)=E\,\psi(x)
\end{equation}
if and only if
\begin{equation}
   \label{phi}
\ph(x)=\psi(ix)
\end{equation}
is a solution of
\begin{equation}
  \label{hu}
\left[-\d_x^2+U(x)\right]\,\ph(x)=-E\,\ph(x)\,.
\end{equation}
Just as in the hyperbolic case, we see by inspection that the
trigonometric Razavy potential can be expressed as an algebraic QES
potential in two different ways. Indeed, the potential \eqref{tr} is
a particular case of two entries in the table of periodic
one-dimensional QES potentials given in Ref.~\cite{GKOqes94}: case 4,
\begin{equation}
  U(x)=A\,\sin^2\sqrt\nu x+B\,\sin\sqrt\nu x+C\,\tan\sqrt\nu x\,\sec\sqrt\nu
x
  +D\,\sec^2\sqrt\nu x
  \label{case4}\,
\end{equation}
for
\begin{equation}
  \label{c4}
\nu=4\,,\qquad A=-\z^2\,,\qquad B=2\,M\,\z\,,\qquad C=D=0
\end{equation}
(after performing the translation $x\mapsto x-\pi/4$; notice that
one-dimensional QES potentials where classified in
Ref.~\cite{GKOqes94} only up to an arbitrary translation), and case 5,
\begin{equation}
  U(x)=A\,\cos 4\sqrt\nu x + B\,\cos 2\sqrt\nu x + C\,\sin 2\sqrt\nu x
  +D\,\sin 4\sqrt\nu x
  \label{case5}
\end{equation}
with
\begin{equation}
  \label{c5}
\nu=1\,,\qquad A=-\frac{\z^2}2\,,\qquad B=2\,M\,\z\,,\qquad C=D=0.
\end{equation}
The four parameters $A,B,C,D$ appearing in the table of periodic QES
potentials of Ref.~\cite{GKOqes94} are not independent, but must
satisfy a single algebraic constraint; it can be verified that the
choices of parameters \eqref{c4} and \eqref{c5} do indeed satisfy this
constraint. Let us also note at this point that, although the
representation \eqref{case4}--\eqref{c4} was known to Turbiner, \cite{Tu88},
and Shifman, \cite{Sh89}, the second representation appears to be new.

Let us now construct the systems of weakly orthogonal polynomials
associated to the representations \eqref{case4} and \eqref{case5} of the
trigonometric Razavy potential as a QES potential.

Consider, in the first place, the representation \eqref{case4}. Instead
of proceeding directly, along the lines sketched in the previous
section for the hyperbolic case, we shall exploit the fact that the
trigonometric and the hyperbolic Razavy potentials are related by the
anti-isospectral transformation \eqref{hv}--\eqref{hu}. We saw in the
previous section (Eq.~\eqref{Hgauge}) that
$$
-\d_x^2+V(x)=
\left.\hat\mu(z)\cdot\Hhg(z)\cdot\frac1{\hat\mu(z)}\right|_{z=\cosh
2x}\,;
$$
performing the change of independent variable $x\mapsto i x$ we obtain
$$
\d_x^2+V(ix)=
\left.\hat\mu(z)\cdot\Hhg(z)\cdot\frac1{\hat\mu(z)}\right|_{z=\cos
2x}
$$
or, equivalently,
$$
-\d_x^2+U(x)=
-\left.\hat\mu(z)\cdot\Hhg(z)\cdot\frac1{\hat\mu(z)}\right|_{z=\cos
2x}.
$$
Thus, the gauge Hamiltonian $\Hhgp$ associated to the potential $U(x)$
in this case is simply
\begin{equation}
\Hhgp(z)=-\Hhg(z)\,.
  \label{Hhgp}
\end{equation}
Since (cf.~Ref.~\cite{FGRorth96}, Eq.~[41]) the coefficients $a_k$ and
$b_k$ defining through Eq.~\eqref{rr} the orthogonal polynomial system
associated to a QES potential are, respectively, quadratic and linear
in the gauge Hamiltonian, it follows that the recurrence relation
satisfied by the orthogonal polynomial system
$\left\{\Pp_k(E)\right\}_{k=0}^\infty$ associated to the
representation \eqref{case4}--\eqref{c4} is
$$
\Pp_{k+1}(E) = (E+b_k)\,\Pp_k(E)-a_k\,\Pp_{k-1}(E),\qquad k\geq 0\,.
$$
Comparing with \eqref{rr} we immediately obtain the relation
\begin{equation}
  \Pp_k(E)=(-1)^k \hat P_k(-E)\,,\qquad k\ge 0\,.
  \label{Pp}
\end{equation}
It can be easily verified through a routine calculation similar to the
one performed in the previous section that the general procedure
described in Ref.~\cite{FGRorth96} to construct the orthogonal
polynomial system associated to a QES potential, when applied to the
representation \eqref{case4} of the potential \eqref{tr}, does indeed
yield the result \eqref{Pp}.

Let us now turn to the representation \eqref{case5}. We cannot
directly apply the previous reasoning in this case, since the
composition of the change of coordinate $z=e^{2x}$ with the
anti-isospectral mapping
$x\mapsto ix$ leads to the complex change of variable $z=e^{2ix}$, while
the correct one for this case is (cf.~\cite{GKOqes94})
\begin{equation}
    z = \tan x\,.
    \label{zp}
\end{equation}
It is therefore easier to apply, as in the previous section, the
general procedure described in \cite{FGRorth96}. Using the techniques
explained in Ref.~\cite{GKOnorm93}, we readily find the following
expression for the coefficients $A,B,C,D$ in \eqref{case5} in terms of
four independent parameters $\bh,\ch,\dh\in\R$ and
$n\in\N\union\{0\}$:
\begin{gather}
    A=\frac1{32\,\nu}(\bh-\ch-\dh)(\bh+\ch-\dh)\,,
    \qquad B=\frac1{8 \nu}\left[\dh^2-\bh^2+4\nu\ch(n+1)\right],\notag\\
    C=\frac1{8\nu}\left[\ch(\bh+\dh)+4\nu(\bh-\dh)(n+1)\right],
    \qquad D=\frac\ch{16\,\nu}\,(\dh-\bh)\,.
    \label{pparam}
\end{gather}
Comparing \eqref{c5} with \eqref{pparam} we readily obtain
\begin{equation}
    \bh=\dh=0\,,\qquad
    \ch=4\z\,,\qquad n=M-1\,.
    \label{zm}
\end{equation}
From \eqref{zm} it follows (cf.~Ref.~\cite{GKOnorm93}) that the change
of variable \eqref{zp} and the gauge transformation determined by
\begin{equation}
    \tilde\mu^{\text{per}}(z)=(z^2+1)^{\frac{1-M}2}\,
    e^{-\frac\z{z^2+1}}
    \label{mup}
\end{equation}
map the trigonometric Razavy Hamiltonian into the gauge
Hamiltonian
\begin{equation}
    \Htgp(z)=-\left[J_+^2+2J_0^2+J_-^2+4\z\,J_0+
    \frac12(M^2+2\z^2+1)\right]\,,
    \label{htgp}
\end{equation}
in the sense that
\begin{equation}
-\d_x^2+U(x)=\left.\tilde\mu^{\text{per}}(z)\cdot\Htgp(z)\cdot
\frac1{\tilde\mu^{\text{per}}(z)}\right|_{z=\tan x}.
\label{ug}
\end{equation}
As in the previous section, the differential operators
$J_\ep$ ($\ep=0,\pm$) appearing in \eqref{htgp} are defined by
\eqref{Js}, with $n=M-1$. The orthogonal
polynomial system $\left\{\Ptp_k(E)\right\}_{k=0}^\infty$ for this
case is generated by expanding an arbitrary solution $\chitp(z)$ of the
gauged equation
\begin{equation}
    \left[\Htgp(z)-E\right]\chitp(z)=0
    \label{gaugep}
\end{equation}
in the formal power series (cf.~Ref.~\cite{FGRorth96}, Eqs.~(29) and (40))
\begin{equation}
    \chitp(z)=(z+i)^{M-1}\sum_{k=0}^\infty \frac{(-1)^k}{(2\z)^k
k!}\,\Ptp_k(E)\,\left(\frac{z-i}{z+i}\right)^k.
    \label{chitp}
\end{equation}
From \eqref{ug} and \eqref{gaugep} it follows that
\begin{equation}
    \ptp(x)=\left.\tilde\mu^{\text{per}}(z)\,\chitp(z)\vrule
     depth 5pt width0pt\right|_{z=\tan x}
    \label{psip}
\end{equation}
is a formal solution of the Schr\"odinger equation
\begin{equation}
    \left[-\d_x^2+U(x)\right]\ptp(x)=E\,\ptp(x)\,.
    \label{schp}
\end{equation}
Applying the change of variables $x\mapsto -ix$ we deduce that
$\ptpm(-ix)$ satisfies the dual
equation
\begin{equation}
    \left[-\d_x^2+V(x)\right]\ptpm(-ix)=E\,\ptpm(-ix)\,.
    \label{dual}
\end{equation}
On the other hand, a direct calculation shows that
\begin{equation}
    \ptpm(-ix)=\left.i^{M-1}e^{-\frac\zeta2}\,\tilde\mu(z)\,
    \sum_{k=0}^\infty\frac{\Ptp_k(-E)}{(2\z)^k
    k!}\,z^k\right|_{z=e^{2x}},
    \label{psim}
\end{equation}
where $\tilde\mu(z)$ is defined by \eqref{mu3}. From \eqref{dual}
and \eqref{psim} it
follows that the formal power series
\begin{equation}
\sum_{k=0}^\infty\frac{\Ptp_k(-E)}{(2\z)^k
k!}\,z^k
\label{chitpm}
\end{equation}
must be proportional to the function $\tilde\chi_E(z)$ generating the
orthogonal polynomial system associated to the the representation
\eqref{case3}--\eqref{param3} of the hyperbolic Razavy potential as an
algebraic QES potential (see Eq.~\eqref{series3}). Since both power
series \eqref{chitpm} and
\eqref{series3} have constant term equal to one, they
must be equal. Equating their coefficients we obtain the relation
\begin{equation}
    \Ptp_k(E)=(-1)^k \tilde P_k(-E)\,,\qquad k\ge 0\,.
    \label{Ptp}
\end{equation}
\subsection{The band spectrum}
\label{sub.band}
The results of the previous subsection imply that, as in the
hyperbolic case, the trigonometric Razavy potential is algebraically
QES when $M\in\N$ (see, for instance, Eq.~\eqref{zm}). We shall
see in this subsection how this fact can be used to exactly compute a
certain number of gaps in the band energy spectrum of the potential
\eqref{tr}, and shall furthermore determine the location of these gaps in
the spectrum.

We start by briefly recalling certain well-known facts about the
energy spectrum of a Schr\"odinger operator
\begin{equation}
    H=-\d_x^2+U(x)\,,
    \label{Hp}
\end{equation}
where $U$ is a continuous periodic function of period $a>0$:
\begin{equation}
    U(x)=U(x+a)\,,\qquad\forall x\in\R\,.
    \label{Up}
\end{equation}
A real number $E$ belongs to the spectrum of $H$ if the
differential equation
\begin{equation}
    (H-E)\,\psi = 0
    \label{eigp}
\end{equation}
has a bounded nonzero solution $\psi(x)$. It can be shown,
\cite{Ho86, RS78}, that the spectrum of the operator
\eqref{Hp}--\eqref{Up}
is the union of an infinite number of closed intervals ({\bf energy
bands}) $[E_0,\bar E_1]$, $[\bar E_2,E_1]$, $[E_2,\bar E_3],\dots$, where
$$
E_0<\bar E_1\le\bar  E_2<E_1\le E_2<\bar E_3\le \bar E_4<\dots
$$
and
$$
\lim_{k\to\infty}E_k=\lim_{k\to\infty}\bar E_k=+\infty\,.
$$
The {\bf gaps} in the energy spectrum are thus the (possibly empty)
open intervals
$$
(\bar E_1,\bar E_2)\,,\;(E_1,E_2)\,,\;\dots\;,\; (\bar
E_k,\bar E_{k+1})\,,\; (E_k,E_{k+1})\,,\;\dots\;\,;\qquad
k=1,2,\dots\,.
$$
The numbers $E_k$ ($k=0,1,2,\dots$) are characterized by the
existence of a nonzero $a$-periodic solution $\psi_k$ of the
differential equation \eqref{eigp} for $E=E_k$. The latter condition is
clearly equivalent to $\psi_k$ being a nonzero solution
of the Sturm--Liouville problem with periodic boundary conditions
\begin{align}
    &(H-E)\,\psi(x)=0\,,\qquad 0<x<a\,;\notag\\
    &\psi(0)=\psi(a)\,,\quad\psi'(0)=\psi'(a)\,,
    \label{Eper}
\end{align}
with eigenvalue $E=E_k$. Note that $E_k=E_{k+1}$ if and only if
\eqref{Eper} has two linearly independent solutions. Likewise, the
numbers $\bar E_k$ ($k=1,2,\dots$) are characterized by the fact that
the differential equation \eqref{eigp} with $E=\bar E_k$ possesses a
nonzero anti-periodic ($2a$-periodic) solution $\pb_k$, that is, a
solution $\pb_k$ such that
$$
\pb_k(x+a)=-\pb_k(x)\,,\qquad\forall x\in\R\,.
$$
Equivalently, $\pb_k$ is a solution of the Sturm--Liouville problem
\begin{align}
    &(H-E)\,\psi(x)=0\,,\qquad 0<x<a\,;\notag\\
    &\psi(0)=-\psi(a)\,,\quad\psi'(0)=-\psi'(a)
    \label{Eaper}
\end{align}
with eigenvalue $E=\bar E_k$. Finally, it is shown in
Ref.~\cite{In56} that for $k=0,1,2,\dots$ the eigenfunction $\psi_k$
has exactly $k+\pi(k)$ zeros in the interval $[0,a)$, where
$$
\pi(k)=\frac12\left[1+(-1)^{k+1}\right]
$$
is the parity of $k$. A straightforward adaptation of Ince's proof
shows that $\pb_k$ has exactly $k+\pi(k)-1$ zeros in $[0,a)$, where
now $k=1,2,\dots$.

\begin{figure}[H]
    \centering
\epsfig{file=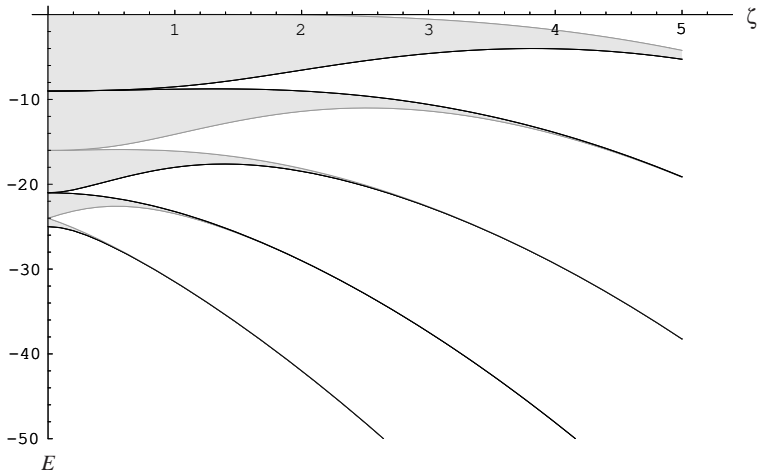, width=10cm}
\begin{quote}
\caption{Allowed energy bands (in gray) for the
potential \eqref{tr} with $M=5$ as a function of $\z$. The solid
boundary lines correspond to the algebraically computable (periodic)
energies, while the gray boundary lines have been obtained
numerically.}\label{fig.bandas}
\end{quote}

\end{figure}

Let us now turn to the trigonometric Razavy potential \eqref{tr},
which we know from the previous discussion to be algebraically QES for
$M\in\N$. For convenience, we shall use in what follows the
representation \eqref{case5}--\eqref{c5} of the trigonometric Razavy
potential as an algebraic QES potential. From \eqref{Ptp} it follows
that the polynomials $\Ptp_k(E)$
satisfy the recurrence relation
$$
\Ptp_{k+1}(E)=(E+b_k)\, \Ptp_k(E)-a_k\, \Ptp_{k-1}(E)\,,\qquad
k\ge0\,,
$$
where the coefficients $a_k$ and $b_k$ are given by \eqref{ab3} with
$n=M-1$. In particular, the coefficient $a_M$ vanishes identically,
which in turn implies that if $\ep_j$ is a root of the polynomial
$\Ptp_M(E)$ then the series \eqref{chitp} truncates at $k=M-1$, and
therefore the function $\pp_{\ep_j}(x)$ given by \eqref{psip} is a
regular, bounded solution of the Schr\"odinger equation \eqref{schp}. Since
it
can be shown \cite{FGRorth96} that the polynomial $\Ptp_M$ has exactly
$M$ different real roots, we can algebraically compute
$M$ solutions of the Schr\"odinger equation
\eqref{schp} of the form (cf.~Eqs.~\eqref{mup}, \eqref{chitp}, and
\eqref{psip})
\begin{equation}
    \pp_{\ep_j}(x)=e^{-\frac\z2\cos
    2x-i(M-1)x}\,\varphi_j\left(e^{2ix}\right)\,,\qquad1\le j\le M\,,
    \label{algsol}
\end{equation}
where $\ep_j$ is any of the roots of $\Ptp_M(E)$ and $\varphi_j$ is a
polynomial of degree at most $M-1$. From the latter
equation we find that
\begin{equation}
    \pp_{\ep_j}(x+\pi)=(-1)^{M-1}\pp_{\ep_j}(x)\,,
    \label{period}
\end{equation}
and therefore the $M$ algebraically computable solutions
\eqref{algsol} are $\pi$-periodic for $M$ odd, and anti-periodic for
$M$ even. Furthermore, from \eqref{algsol} it also follows that each
of the $M$ algebraic eigenfunctions $\pp_{\ep_j}$ has at most $M-1$
roots in the interval $[0,\pi)$. Thus for $M$ even the algebraic
eigenfunctions \eqref{algsol} coincide with the $M$ lowest
anti-periodic eigenfunctions $\pb_k(x)$ ($k=1,\dots,M$), and the
algebraically computable energies --- the zeros $\ep_j$ of the
critical polynomial $\Ptp_M(E)$ --- are the lowest $M$
``anti-periodic'' energies $\bar E_1,\dots,\bar E_M$. Likewise, for
$M$ odd the algebraic eigenfunctions are the lowest $M$ periodic
eigenfunctions $\psi_k(x)$ ($k=1,\dots,M$), and the corresponding
algebraic energies are the $M$ lowest ``periodic'' energies
$E_0,\dots,E_{M-1}$. Since the gaps in the energy spectrum of a
periodic Hamiltonian are limited by two energies of the \emph{same}
type (periodic or anti-periodic), this means that the knowledge of the
algebraic energies allows us to exactly compute a certain number of
gaps in the spectrum. More precisely, when $M$ is
even then we can algebraically compute the first $M/2$ anti-periodic
gaps $(\bar E_1,\bar E_2),(\bar E_3,\bar E_4),\dots,(\bar E_{M-1},\bar
E_M)$ in the energy spectrum of the trigonometric the Razavy
Hamiltonian or, equivalently, the first $M/2$ \emph{odd} gaps.
Similarly, when $M$ is odd then the algebraically computable energies
are the ground state $E_0$ and the first $(M-1)/2=[M/2]$ periodic gaps
$(E_1,E_2),(E_3,E_4),\dots,(E_{M-2}, E_{M-1})$ or, what is the same,
the ground state and the first $[M/2]$ \emph{even} gaps in the energy
spectrum. In particular, for $M$ even we can algebraically
compute the \emph{first} gap $(\bar E_1,\bar E_2)$ in the energy
spectrum of the trigonometric Razavy Hamiltonian, while for $M$ odd
we can always compute the ground state $E_0$. Note also that,
since the algebraically computable gaps are never consecutive, we
cannot exactly compute any of the allowed energy bands.
Figure~\ref{fig.bandas} shows the first five allowed
energy bands for the trigonometric Razavy potential as a
function of the parameter $\z$ for $M=5$.

The differential equation \eqref{eigp} with the potential \eqref{tr}
is well-known in the theory of periodic differential equations under
the name of {\em Whittaker--Hill's equation,} or Hill's three-term
equation, \cite{Ar64, MW79}. It is of interest in the latter
context mainly because, unlike the much better known Mathieu's equation,
for certain values of the spectral parameter $E$ it admits so called
{\bf finite solutions}, i.e., solutions of the form
$$
\psi(x)=e^{-\frac\z2\cos 2x}\varphi(x)\,,
$$
where $\varphi(x)$ is a trigonometric polynomial. It follows from
Eq.~\eqref{algsol} that the algebraic eigenfunctions obtained in this
section are finite solutions. In fact, the converse also holds,
namely \emph{all} finite solutions are algebraic eigenfunctions. This
follows at once from Theorem 7.9 of Ref.~\cite{MW79}, which states (in
our notation) that for each $M\in\N$ the Whittaker--Hill equation has
at most $[M/2]$ gaps of periodic (if $M$ is odd) or anti-periodic (if
$M$ is even) type. For even $M$, we have shown that there are exactly
$M$ anti-periodic algebraic eigenfunctions, whose corresponding $M$
eigenvalues define \emph{exactly} $M/2$ gaps of anti-periodic type.
Since, for these values of $M$, there are also exactly $M$
anti-periodic finite solutions and eigenvalues, \cite{Ar64}, it
follows from the Theorem quoted above that the finite solutions
coincide (up to a constant factor) with the algebraic eigenfunctions.
A similar argument is valid when $M$ is odd.

In order to compare our findings with the classical theory, it is more
convenient to use the representation \eqref{case4}--\eqref{c4}.
Applying the anti-isospectral transformation to Eqs.~\eqref{mu2} and
\eqref{series2} of Section \ref{sec.razavy}, it follows that the
algebraic (unnormalized) eigenfunctions can be classified as follows:
\begin{align}
    \intertext{\noindent $M$ even:}
    \psi(x)&=e^{-\frac\z2\cos 2x}\,\sum_{k=0}^{\frac M2-1}\frac{\hat
    P_k^{0+}(-E)}{(2k+1)!}\,\cos^{2k+1}x\,,
    \label{ee}\\
     \psi(x)&=e^{-\frac\z2\cos 2x}\,\sin x\,\sum_{k=0}^{\frac
     M2-1}\frac{\hat P_k^{0-}(-E)}{(2k)!}\,
     \cos^{2k}x\,; \label{eo}\\
    \intertext{\noindent $M$ odd:}
        \psi(x)&=e^{-\frac\z2\cos 2x}\,\sum_{k=0}^{\frac{M-1}2}\frac{\hat
    P_k^{+0}(-E)}{(2k)!}\,\cos^{2k}x\,,
    \label{oe}\\
     \psi(x)&=e^{-\frac\z2\cos 2x}\,\sin
     x\,\sum_{k=0}^{\frac{M-3}2}\frac{\hat
     P_k^{-0}(-E)}{(2k+1)!}\,\cos^{2k+1}x\,.
    \label{oo}
\end{align}
In the above formulae, the polynomials $\hat P^{\sigma\eta}_k$ are
defined by the recursion relation \eqref{rr}--\eqref{abKM}, and $E$ is
one of the algebraically computable energies, i.e, $-E$ is a root of
the critical polynomials $\hat P^{0+}_{M/2}$, $\hat P^{0-}_{M/2}$,
$\hat P^{+0}_{(M+1)/2}$, and $\hat P^{-0}_{(M-1)/2}$, respectively.
Comparing with the formulae in Section 7.4.1 of Ref.~\cite{Ar64} we
easily find that, if $\psi(x)$ is an algebraic eigenfunction of one of
the four types \eqref{ee}--\eqref{oo}, then $e^{\frac\z2\cos 2x}\psi(x)$ is
proportional, respectively, to the {\bf Ince polynomial} $C_{M-1}^{2k+1},\;
S_{M-1}^{2k+1},\;C_{M-1}^{2k}$, and $S_{M-1}^{2k}$, where (in the
notation of Ince, cf.~Ref.~\cite{Ar64}) $E=a_{M-1}^{2k+1},\;
b_{M-1}^{2k+1},\;a_{M-1}^{2k}$, or $b_{M-1}^{2k}$,
respectively.

The results of this section can therefore be interpreted as providing
a deep Lie-algebraic justification for the exceptional fact that the
Whittaker--Hill equation admits finite solutions.  This observation is
further corroborated by the fact that other periodic Schr\"odinger
equations known to have finite solutions (in a slightly more general
sense) as, for instance, the Lam\'e equation, are also algebraically
QES, \cite{AGI83, Tu89, UZ92, GKOqes94}.  The
above results underscore the close connection between the existence of
finite solutions of Hill's equation, on the one hand, and the
algebraic QES character of its potential, on the other. This
remarkable connection certainly deserves further investigation.

\end{document}